\documentclass[10pt,letterpaper]{article}
\usepackage{opex3}
\usepackage{color}

\usepackage{amsmath,amssymb,graphicx}
\usepackage[english]{babel}
\usepackage{romannum}
\usepackage{tabu}
\usepackage{cite}
\usepackage{multirow}
\usepackage[usenames, dvipsnames]{xcolor}

\RequirePackage{tikz}

\newcommand{\paren}[1]{\left({#1}\right)}
\newcommand{\sqpr}[1]{\left[{#1}\right]}
\newcommand{\dydx}[2]{\frac{\mathrm{d} #1}{\mathrm{d} #2}}

\newcommand{\sech}{\mathrm{sech}}

\newcommand{\dint}{\mathrm{d}}
\newcommand{\dt}{\mathrm{d}t}
\newcommand{\pypx}[2]{\frac{\partial #1}{\partial #2}}
\newcommand{\pynpx}[3]{\frac{\partial^{#3} #1}{\partial {#2}^{#3}}}

\renewcommand{\vec}[1]{\mathbf{#1}}

\renewcommand{\frac}[2]{\dfrac{#1}{#2}}

\newcommand{\abs}[1]{\left|{#1}\right|}

\newcommand{\quoeq}[1]{Eq.~(\ref{#1})}

\begin{document}

\title{Comparison of Models of Fast Saturable Absorption in Passively Modelocked Lasers}

\author{Shaokang Wang$^{*}$, Brian S. Marks, and Curtis R. Menyuk}

\address{
Department of Computer Science and Electrical Engineering, \\
University of Maryland, Baltimore County,  1000 Hilltop Circle, Baltimore, MD, 21250 
}

\email{swan1@umbc.edu}

\begin{abstract}
Fast saturable absorbers (FSAs) play a critical role in stabilizing many passively modelocked lasers. 
The most commonly used averaged model to study these lasers is the Haus modelocking equation (HME) that includes a third-order nonlinear FSA.
However, it predicts a narrow region of stability that is inconsistent with experiments. 
To better replicate the laser physics, averaged laser models that include FSAs with higher-than-third-order nonlinearities have been introduced. 
Here, we compare three common FSA models to each other and to the HME using the recently-developed boundary tracking algorithms. 
The three FSA models are the cubic-quintic model, the sinusoidal model, and the algebraic model. 
We find that all three models predict the existence of a stable high-energy solution that is not present in the HME and have a much larger stable operating region. 
We also find that all three models predict qualitatively similar stability diagrams. 
We conclude that averaged laser models that include FSAs with higher-than-third-order nonlinearity should be used when studying the stability of passively modelocked lasers. 
\end{abstract}


\bibliographystyle{osajnl}	
\bibliography{myrefs}		

\section{Introduction}

Over the past few decades, ultra-short optical pulses that are produced by passively modelocked lasers have been used in a broad range of fields~\cite{Diddams:10,kartner2014few}.  
Saturable absorption plays a critical role in stabilizing in these lasers and determining their pulse shape~\cite{haus1975-2,Haus1975}.  
There has been considerable controversy over how best to model the saturable absorber~\cite{Paschotta:2004,kutz:06,Ablowitz:11,Soto-Crespo:96,Chen:92,leblond2002, komarov2005, kamarov-quintic2005}, and resolution of this issue has been hampered until recently by lack of a theoretical tool that is capable of accurately determining the stability boundaries in the parameter space for the different models of saturable absorption.  

The most commonly used averaged model of passively modelocked lasers with a fast saturable absorber (FSA) and a slow saturable gain is the Haus modelocking equation (HME)~\cite{haus902165}. This model has been used to study solid-state lasers~\cite{Haus1975,Haus:1991,Dunlop:98}, fiber lasers~\cite{Chen:92,Hofer1992,kodama1992soliton,newbury2005}, as well as semiconductor lasers~\cite{kartner1996}. 
The HME may be written as~\cite{haus:3049}
\begin{equation}\label{eq:hme}
\begin{aligned}\displaystyle
\pypx{u}{z}=\bigg[-i\phi-\frac{l}{2}+\frac{g(\abs{u})}{2} \bigg(1+\frac{1}{2\omega_g^2}\pynpx{}{t}{2}\bigg) - \frac{i\beta''}{2}\pynpx{}{t}{2} + i\gamma|u|^2 \bigg]u + f_{\mathrm{sa}}(|u|)u, 
\end{aligned}
\end{equation}
where $u(t,z)$ is the complex field envelope, $t$ is the retarded time, $z$ is the propagation distance, $\phi$ is the phase rotation per unit length, $l$ is the linear loss coefficient, $g(|u|)$ is the saturated gain, $\beta''$ is the group velocity dispersion coefficient, $\gamma$ is the Kerr coefficient, $\omega_g$ is the gain bandwidth, and $f_{\mathrm{sa}}(|u|)$ is the fast saturable absorption. 
It is common in studies of the HME to set $\dint\phi/\dint z=0$~\cite{kutz:06,haus902165,Haus:1991,Hofer1992,Kapitula:02}, in which case the stationary pulse solution---which we also refer to as the equilibrium solution---rotates at a constant rate as a function of $z$. 
In a stability analysis, it is more convenient to start from a stationary solution of Eq.~(\ref{eq:hme}), in which case $\phi=\phi_0$ must be found along with $u_0(t)$~{\cite{Wang:2014}}, which is what we will do here. 
In the HME, it is assumed that the gain response of the medium is much longer than the roundtrip time $T_R$, in which case the saturable gain becomes
\begin{align}\label{eq:gain_sat}
g(\abs{u})=\frac{g_0}{1+P_\mathrm{av}(\abs{u})/P_\mathrm{sat}},
\end{align}
where $g_0$ is the unsaturated gain, $P_{\mathrm{av}}(\abs{u})$ is the average power, 
and $P_\mathrm{sat}$ is the saturation power. 
We may write $P_\mathrm{av}(\abs{u})=\int_{-T_R/2}^{T_R/2} |u(t,z)|^2 \dt/T_R$.
In this article, we focus on the anomalous chromatic dispersion regime in which $\beta''<0$.
In the HME, the FSA function $f_{\mathrm{sa}}$ has the simple form
\begin{align}\label{eq:cubic}
f_{\mathrm{sa}}(|u|)=\delta|u|^2.
\end{align}

The HME has been successful in exploring many of the qualitative features of passively modelocked lasers. 
However, the predictions of the HME for the instability thresholds are unrealistically pessimistic~\cite{Kapitula:02,Renninger2008,Cundiff2003}. 
One particular reason is that Eq.~(\ref{eq:cubic})---which behaves as an ideal saturable absorber---provides unlimited gain, which grows cubically as the input pulse amplitude grows. 
This mechanism in theory leads to an infinite growth of the pulse energy and hence a saddle-node instability when the cubic coefficient $\delta$ becomes sufficiently large~\cite{Kapitula:02}. 

In order to better replicate the laser physics, different models of the FSA have been developed.
However, there has been controversy concerning which FSA model is ``best'' to use in the sense that it best reproduces the broad stability region that is observed in experiments~\cite{Ablowitz:11,Chong:06,Cundiff2003}. 
In this article, we address this issue using the recently developed boundary-tracking algorithm~\cite{Wang:2014,Wang_IPC2013} to examine three common FSA models, which all have a quintic nonlinearity at second-lowest order in a Taylor expansion of the intensity. 
We show that while these models are quantitatively different, they produce similar stability diagrams in the parameter space.
Consequently, none of these models is to be preferred on the basis of their qualitative features, and only careful comparison with experiments with an appropriate selection of the model parameters can distinguish among them.
However, all three models predict the existence of a stable high-energy solution that is not present in the HME, and, as a consequence, all three models have a much larger stable operating region in the parameter space.

The reminder of this article is organized as follows: 
We introduce and compare an extended modelocking equation with different FSA models in Sec.~\ref{sec:mfsa}.
In Sec.~\ref{sec:stability}, we compare the dynamical structure of the stability regions of the extended model equation with the three different FSA models. 
We show two cases in which singular solutions exist but remain stable. 
In Sec.~\ref{sec:experiments}, we compare the computational pulses with experimental results, and we conclude that modelocking models with FSA models that includes higher-than-third order terms should always be preferred to the HME when studying the stability of passively modelocked lasers with averaged models. 

\section{Models of Fast Saturable Absorption\label{sec:mfsa}}

Two physics-based models of fast saturable absorption (FSA) are commonly used.
The first and oldest of these models is the algebraic model. 
This model has been applied particularly to solid state lasers~\cite{Sander:09}, but also to analyze the noise and stability of passively modelocked lasers in  general~\cite{Paschotta:2004,Ablowitz:11,Horikis:14}. 
In this model, it is assumed that the absorbing medium is a two-level system in which the response time of the medium is fast compared to the pulse duration, so that the population of the upper state is proportional to $|u(t)|^2$. 
In this case, we find~\cite{haus902165,haus:3049,Paschotta:2004,Ablowitz:11,Horikis:14}
\begin{align}\label{eq:two-level}
\pypx{u}{z}\bigg|_\mathrm{ab}=f_\mathrm{ab}(|u|)u = -\frac{f_0 u}{1+|u(t)|^2/P_\mathrm{ab}},
\end{align}
where $\partial u/\partial z|_\mathrm{ab}$ is the contribution to the loss from the absorbing mechanism, $f_0$ is a constant, and $P_\mathrm{ab}$ is the saturation power of the absorber. 
A common way of simplifying the modelocking model is to incorporate the unsaturated loss $s_0$ into the linear loss $l$~\cite{haus902165,haus:3049}, i.e., $l/2=\alpha+f_0$, where $\alpha$ donates the total loss that is not due to the material absorber, such as losses from the end mirrors and couplers. 
We then obtain 
\begin{align}~\label{eq:s_al}
f_\mathrm{sa}(|u|) = -\sqpr{f_\mathrm{ab}(|u|)-f_0} = -\frac{f_0|u|^2/P_\mathrm{ab}}{1+|u|^2/P_\mathrm{ab}}.
\end{align}
When the saturation is relatively weak, i.e., $|u(t)|^2\ll P_\mathrm{ab}$, we find
\begin{align}\label{eq:solid-taylor}
f_\mathrm{ab}(|u|) = \frac{f_0}{P_\mathrm{ab}}|u(t)|^2-\frac{f_0}{P^2_\mathrm{ab}}|u(t)|^4-\cdots.
\end{align}
By keeping only the leading-order term, we can obtain a cubic model, i.e., \quoeq{eq:cubic}, identifying $\delta={f_0}/{P_\mathrm{ab}}$. 
Similarly, by truncating this expansion at the order $|u|^4$, we obtain a cubic-quintic model
\begin{align}\label{eq:s_cq}
f_\mathrm{sa}(|u|) = \delta|u(t)|^2-\sigma|u(t)|^4,
\end{align}
where $\sigma={f_0}/{P^2_\mathrm{ab}}$.

The second physics-based model assumes that the FSA is due to a combination of nonlinear polarization rotation and polarization selective elements that attenuate low intensities more than high intensities. 
In this model, we have~\cite{Chen:92,leblond2002, komarov2005, kamarov-quintic2005}
\begin{align}\label{eq:s_sn}
f_\mathrm{ab}(|u|) = -f_0+f_1\cos\paren{\mu |u|^2-\nu},
\end{align}
where the constants $f_0$, $f_1$, $\mu$, and $\nu$ depend on the settings of the polarization selective elements and the amount of nonlinear polarization rotation in one pass through the laser. 
If we may assume $\mu|u|^2\ll1$, then
\begin{equation}
\begin{aligned}\label{eq:expand-sinusoidal}
f_\mathrm{ab}(|u|)= -f_0+f_1\cos\nu+\mu f_1(\sin\nu) |u|^2  - (\mu^2f_1/2)(\cos\nu)|u|^4-\cdots.
\end{aligned}
\end{equation}
The combination $-f_0+f_1\cos\nu$ may be absorbed into the total linear loss.
We then find that the nonlinear terms in \quoeq{eq:expand-sinusoidal} becomes similar to the Taylor expansion in \quoeq{eq:solid-taylor}. 
We note that this sinusoidal model also applies to lasers that use a nonlinear optical loop mirror (NOLM) as the FSA~\cite{Duling1994}.

In this article, we study quantitatively how the different FSA models in Eqs.~(\ref{eq:s_al}),~(\ref{eq:s_cq}), and~(\ref{eq:s_sn}) affect the stability of the modelocking equation of \quoeq{eq:hme}.
To facilitate quantitative comparison, we match up the Taylor series of these three models through order $|u|^4$ by assigning
\begin{equation}
\begin{aligned}
\delta = & \frac{f_0}{P_\mathrm{ab}} = \mu f_1\sin\nu, \nonumber \\
\sigma = & \frac{f_0}{P_\mathrm{ab}^2} = \frac{\mu^2 f_1\cos\nu}{2},
\end{aligned}
\end{equation}
in which $\nu=\pi/4$ and $\mu=2\sigma/\delta$. 
We can then write the nonlinear gain functions of the three FSA models as
\begin{subequations}\label{eq:absorbers}
\begin{align}
f_{\mathrm{sa,cq}}(|u|) =&\ \delta |u|^2-\sigma |u|^4, \label{eq:cubic-quintic}\\
f_{\mathrm{sa,al}}(|u|) =&\ \displaystyle{\frac{\delta |u|^2}{1+\paren{\sigma/\delta}|u|^2}},  \label{eq:algebraic}\\
 f_{\mathrm{sa,sn}}(|u|) =&\ \displaystyle{\frac{\delta^2}{2\sigma}\bigg[\sqrt{2}\sin\paren{\frac{2\sigma}{\delta}|u|^2+\frac{\pi}{4}}-1\bigg]},  
\end{align}
\end{subequations}
where $\sigma>0$, and the sub-indices ``cq'', ``ag'', and ``sn'' represents the cubic-quintic model, the algebraic model, and the sinusoidal model, respectively. 
We refer to $\delta$ as the cubic coefficient and $\sigma$ as the quintic coefficient. 
Compared with  Eqs.~(\ref{eq:cubic}),~(\ref{eq:s_al}), and~(\ref{eq:s_sn}), we remove the dependence between $\delta$ and $\sigma$ in Eq.~(\ref{eq:absorbers}) to broaden the generality of comparison. 
Here, we refer to \quoeq{eq:hme} with different FSA models as the generalized Haus modelocking equation (GHME).

\begin{figure}[!h]
\begin{center}
\includegraphics[scale=1]{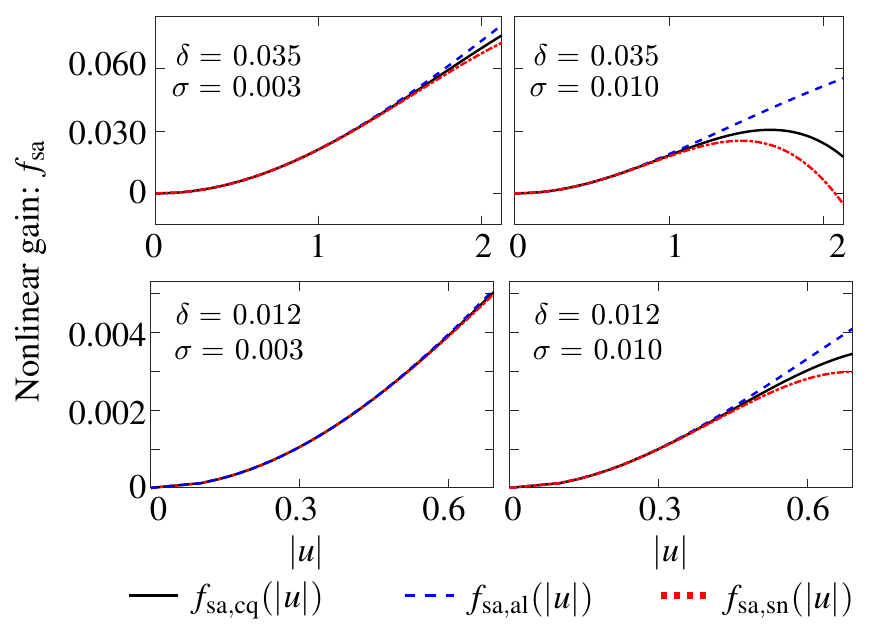} 
\vspace{-0.6cm}
\end{center}
\caption{Comparison of the nonlinear gain for the three models of fast saturable absorption given in Eq.~(\ref{eq:absorbers}).\label{fig:absorbers}}
\vspace{-0.3cm}
\end{figure}

To compare the three FSA models, we show in Fig.~\ref{fig:absorbers} the nonlinear gain $f_\mathrm{sa}(|u|)$ of the three models, in which the input $|u|$ is assumed to be the instantaneous pulse amplitude. 
We show the nonlinear gain with four sets of $(\sigma,\delta)$ values. 
For all three models, we find that the nonlinear gain is almost identical when $|u|$ is small, and becomes increasingly different as $|u|$ grows. 
The difference grows also when $\delta$ and $\sigma$ increase for a given $|u|$.
We also find that the nonlinear gain is the greatest and grows monotonically as $|u|$ increases when using the algebraic model. 
Meanwhile, the nonlinear gain saturates and decreases when $|u|$ becomes sufficiently large with either the cubic-quintic model or the sinusoidal model.
Hence, the algebraic model will lead to greater nonlinear gain and stationary pulses with higher energies than is the case with the other two models. 
The nonlinear gain is smallest with the sinusoidal model. 

\section{Stability of the Generalized Haus Modelocking Equations\label{sec:stability}}

In our analysis, we use the normalization and the parameters for a soliton laser in~\cite{kutz:06,Kapitula:02}. 
The pulse $u$ is normalized with respect to the peak power of the electrical field, the propagation variable $z$ is normalized to the dispersion length, and the retarded time $t$ is normalized to the pulse width. 
Both $u$ and $\phi$ become become unitless with this normalization. 
The set of parameters is listed in Table~\ref{tab:params}. 
(In Table.~1 of~\cite{Wang:2014}, the value of $\omega_g$ was given as $\sqrt{10}/2$ by mistake.)

\begin{table}[!h]
\begin{center}
\begin{tabular}{|l|l|l|l|l|l|l|}
\hline
Parameter& $g_0$ & $l$ & $\gamma$ & $\omega_g$ & $\beta''$ & $T_RP_\mathrm{sat}$\\
\hline
Value & $0.4$ & 0.2 & $4$ & $\sqrt{5}$ & $-2$  & 1 \\
\hline
\end{tabular}
\end{center}
\vspace{-0.3cm}
\caption{Normalized values of parameters.\label{tab:params}}
\vspace{-0.3cm}
\end{table}

\subsection{The cubic-quintic FSA model}

\begin{figure}[!ht]
\begin{center}
\includegraphics[scale=0.9]{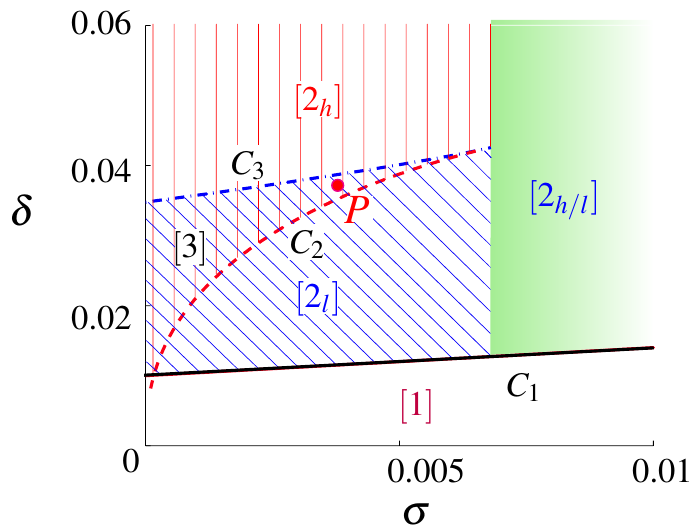} 
\vspace{-0.6cm}
\end{center}
\caption{The stability regions of the GHME with a cubic-quintic saturable absorber $f_\mathrm{sa,cq}(|u|)$. The stability boundaries are marked by three curves, $C_1$, $C_2$, and $C_3$. This figure reproduces Fig.~16 of ref.~\cite{Wang:2014}\label{fig:stability-cqme}}
\end{figure}

In prior work~\cite{Wang:2014,Wang_IPC2013}, we developed boundary tracking algorithms, and we found the stability regions of the GHME with the cubic-quintic model of~\quoeq{eq:cubic-quintic}.
The basic approach of the boundary tracking algorithms is that as parameters vary, we first find a stationary solution $[u_0(t),\phi_0]$ by solving a root-finding problem. 
Then we determine the stability of this stationary solution by calculating the eigenvalues of the linearized evolution equations. 
We show two examples of the distribution of the eigenvalues on the complex plane in Appendix A.
The stationary solution is unstable if any of the eigenvalues have a positive real part. 
We then repeat these computations while tracking the stability boundary in the parameter space. 
In this section, we review the results and show the stability structure in Fig.~\ref{fig:stability-cqme}.
We have discovered a rich dynamical structure.

The $\delta$-axis of Fig.~\ref{fig:stability-cqme} corresponds to the HME case in which the quintic coefficient is zero ($\sigma=0$).
A known analytical solution is
\begin{align}\label{eq:hmesolution}
u_h(t) = A_h\sech^{(1+i\beta_h)}\paren{t/t_h},
\end{align}
where $A_h$, $\beta_h$, and $t_h$ are constants that can be derived from the system parameters.
The stationary solution $u_h(t)$ is stable when $0.01<\delta<0.0348$~\cite{Kapitula:02}. 

When $\sigma\ne0$, the solution in Eq.~(\ref{eq:hmesolution}) does not hold any more, and we have found two stable numerical pulse solutions of the GHME: a low-amplitude solution (LAS) and a high-amplitude solution (HAS). 
The LAS is stable in the blue-hatched region that is marked with $[2_l]$ as shown in Fig.~\ref{fig:stability-cqme}, and it becomes unstable in region $[1]$ (below the curve $C_1$), where the continuous modes become unstable via a Hopf bifurcation or an essential instability~\cite{Kapitula03theevans}.
The amplitude eigenmode becomes unstable when we cross $C_3$ from region $[2_l]$, which corresponds to a saddle-node instability.
Meanwhile, there also exists a second stationary solution~\cite{Wang:2014}, which we refer to as the high-amplitude solution (HAS). 
The HAS is stable in the red-hatched region which is marked with $[2_h]$, and its stability region is lower-bounded by $C_2$, below which the amplitude eigenmode becomes unstable via a saddle-node bifurcation.  
The LAS and the HAS coexist and remain stable in region [3] which is bounded by $C_2$ and $C_3$.
A region like that of region [3] in which the LAS and the HAS coexist has recently been experimentally confirmed~\cite{Bao2015}.
The LAS and the HAS merge into one single solution in region $[2_{h/l}]$ in which it remains stable. 
The HAS does not exist for the HME, which is the reason the behavior of the GHME is qualitatively different from the HME. 
We also find that the HAS remains stable until $\delta$ increases up to $\sim$9.51, which is almost a factor of 280 greater than the HME's stability limit~\cite{Wang:2014}.
For a given $\delta$, the HAS also remains stable for any value of $\sigma$ as long as $\sigma>0$, although the stable equilibrium pulse becomes increasingly peaked and narrow as $\sigma$ approaches zero~\cite{Wang:2016-PRA}. 
We summarize which solution becomes unstable on the curves $C_1$, $C_2$, and $C_3$ and the instability mechanism in Table~\ref{tab:mechanisms}.

\begin{table}[!h]
\begin{center}
\begin{tabular}{|c|c|l|}
\hline
Curve&Solution &Instability Mechanism\\
\hline
$C_1$ & LAS & Essential  \\
$C_2$ & LAS & Saddle-node \\
$C_3$ & HAS & Saddle-node \\
\hline
\end{tabular}
\end{center}
\vspace{-0.3cm}
\caption{Instability mechanisms of the GHME shown in Fig.~\ref{fig:stability-cqme}, where LAS represents the low-amplitude solution, and HAS represents the high-amplitude solution. \label{tab:mechanisms}}
\end{table}

\begin{figure}[t]
\begin{center}
\includegraphics[scale=1]{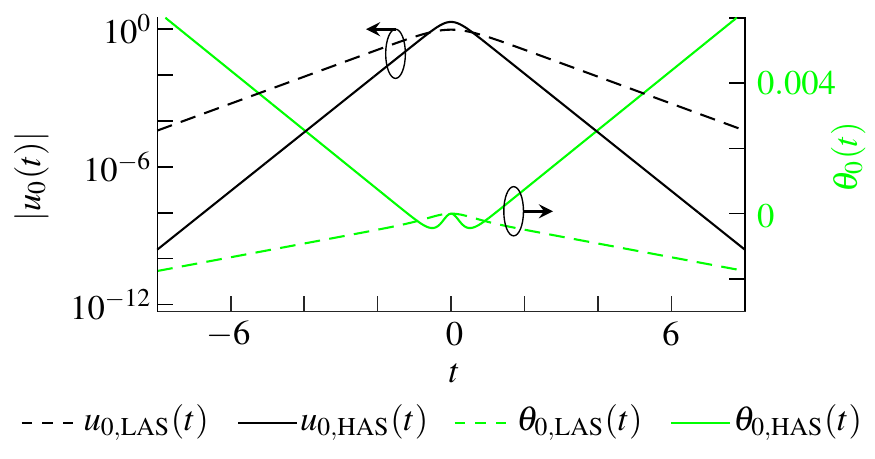} 
\vspace{-0.6cm}
\end{center}
\caption{For the GHME with the cubic-quintic model of \quoeq{eq:cubic-quintic}, the pulse profiles of the low-amplitude solution (LAS) and the high-amplitude solution (HAS) with $(\sigma,\delta)=(0.004,0.036)$. \label{fig:HAS-cq}}
\vspace{-0.3cm}
\end{figure}

In Fig.~\ref{fig:HAS-cq}, we show an example of the pulse profiles of both the LAS and the HAS when $\sigma=0.004$ and $\delta=0.036$, which is at the point $P$ in Fig.~\ref{fig:stability-cqme}. 
We write the stationary pulse profile as 
\begin{align}
u_0(t) = |u_0(t)|\exp\sqpr{\theta_0(t)},
\end{align}
where $\theta_0(t)$ is the phase across the pulse. 
The pulse envelopes of both the LAS and the HAS have a nearly hyperbolic secant profile, in which the amplitude decays exponentially as $t\to\pm\infty$. 
Compared to the LAS, the amplitude of the HAS is visibly higher, and the pulse width is smaller. 
In addition, the HAS has a stronger chirp than does the LAS. 

\subsection{Comparison of stability with different FSA models}

\begin{figure}[!h]
\begin{center}
\includegraphics[scale=0.9]{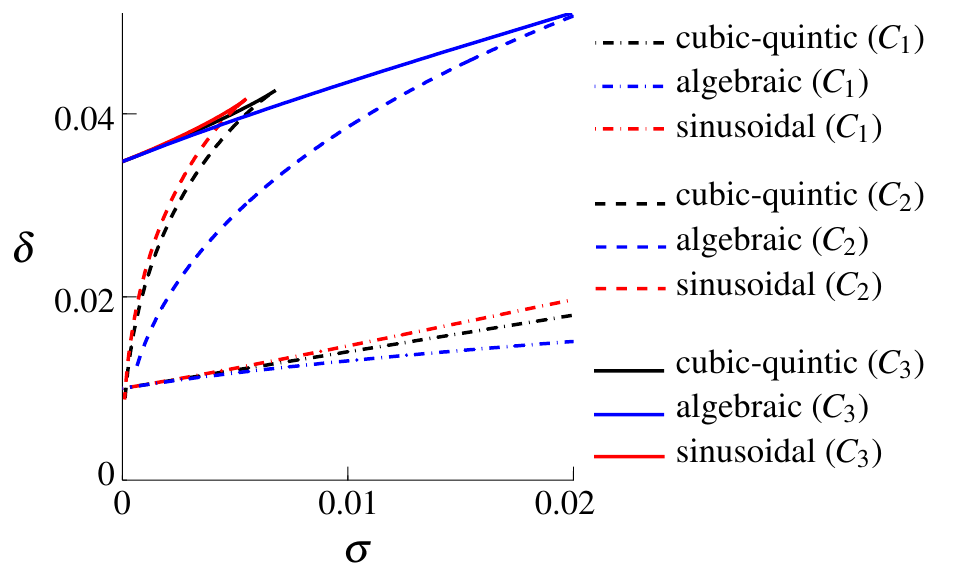} 
\vspace{-0.3cm}
\caption{Stability boundaries for models with the three different absorbers; black (cubic-quintic), blue (algebraic), red (sinusoidal). \label{fig:allmodels}}
\end{center}
\vspace{-0.3cm}
\end{figure}

In Fig.~\ref{fig:allmodels}, we compare regions of stability for the three different FSA models.
We find that the dynamical structures are qualitatively similar for all three models. 
For regions near the origin, all three models feature the three characteristic curves $C_1$, $C_2$, and $C_3$ described in Table~\ref{tab:mechanisms}. 

We first compare the location of $C_1$, which marks the lower boundary where the LAS becomes unstable due to the continuous modes.
When the nonlinear gain in the GHME increases, the energy of the stationary pulse solution grows, and the linear gain $g(|u|)$ becomes increasingly saturated.
Since the stability condition for continuous waves is $[g(|u|)-l]<0$~\cite{Kapitula:02,Wang:2014}, the continuous wave perturbations become increasingly stable.
Hence, a large nonlinear gain in the FSA model implies a large stability region. 
We recall that for fixed values of $(\sigma,\delta)$, the nonlinear gain is largest for the algebraic model and smallest for the sinusoidal model.
This behavior is consistent with what we show in Fig.~\ref{fig:allmodels}. 
With the same value of $\sigma$, the value of $\delta$ on $C_1$ is smallest for the algebraic model, and then is larger for the cubic-quintic model, and is largest for the sinusoidal model.  

Next we analyze the stability limits of the saddle-node instability that leads to an amplitude explosion~\cite{Kapitula:02,Wang:2014}. 
The saddle-node instability occurs when the nonlinear gain becomes so large that it cannot be compensated by the nonlinear loss. 
Thus, with the same values of $\sigma$ and $\delta$, an FSA model that provides a relatively small nonlinear gain tends to be more stable against the occurrence of the saddle-node instability. 
In Fig.~\ref{fig:allmodels}, the onset of the saddle-node instability of the LAS is marked by $C_3$. 
We observe that with the same value of $\sigma$, the stability limit of $\delta$ is  largest with for sinusoidal model, next largest for the cubic-quintic model, and is  smallest for the algebraic model. 

\begin{figure}[!h]
\begin{center}
\includegraphics[scale=1]{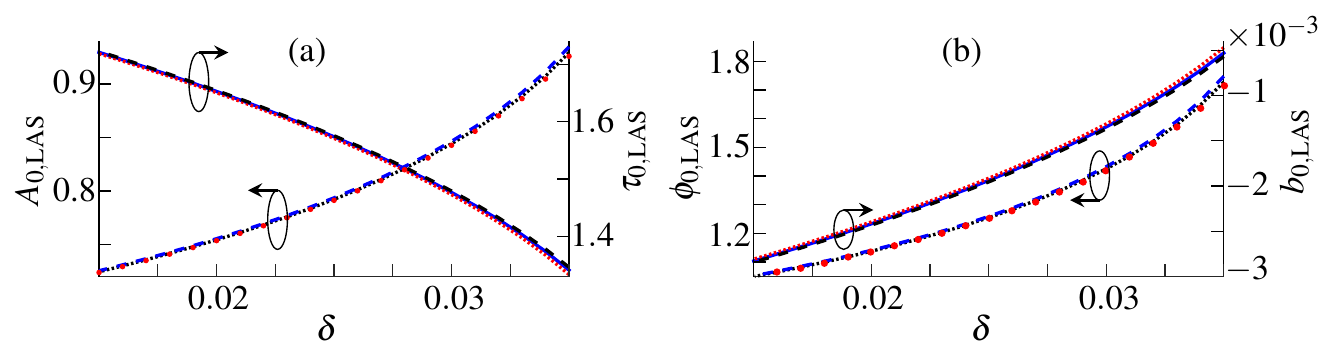} 
\vspace{-1cm}
\end{center}
\vspace{0.1cm}
\caption{Variation of the profile parameters of the low-amplitude solution (LAS) when $\sigma=0.004$. The black curves represent the pulse profiles of the LAS with the cubic-quintic model, and, similarly, the blue curves represent those of the algebraic model, and the red curves represent those of the sinusoidal model.    \label{fig:LAS-profile}}
\vspace{-0.1cm}
\end{figure}

In Fig.~\ref{fig:LAS-profile} we show the pulse profiles of the LAS of all three FSA models when $\sigma=0.004$. 
Here, the parameter $A_0$ is the peak amplitude of the pulse, $\tau_0$ is the FWHM width of the pulse solution, and $b_0$ is the chirp,
\begin{align}
b_0=\mathrm{Im}\frac{\int_{-\infty}^{\infty}t u^*\sqpr{\dint u/\dint t}\dint t} {\int_{-\infty}^{\infty}t^2|u|^2\dint t}.
\end{align} 
We see that the pulse profiles with the three models are nearly identical for the LAS. 
As the cubic gain coefficient $\delta$ increases, we find that the pulse amplitude $A_0$ increases, the pulse width $\tau_0$ decreases, and both the rate of phase rotation $\phi_0$ and the chirp $b_0$ increase. 
Among the three models, the pulse for the algebraic model has the highest amplitude and narrowest width with a fixed ($\sigma,\delta$), while the pulse for the sinusoidal model has the smallest amplitude and the largest pulse width. 
In Fig.~\ref{fig:absorbers}, we see that the curves $C_1$ and $C_3$ are quantitatively close when $0<\sigma<0.006$, which is because the nonlinear gain is nearly identical when $|u|$ is small. 
The small differences in the nonlinear gain of the FSA models as shown in Fig.~\ref{fig:absorbers} are consistent with the small differences in the pulse profiles as shown in Fig.~\ref{fig:LAS-profile}(a), in which the peak amplitude $A_0$ is largest and the pulse width $\tau_0$ is smallest for the algebraic model, while $A_0$ and $\tau_0$ are smallest and largest respectively for the sinusoidal model. 

\begin{figure}[!h]
\begin{center}
\includegraphics[scale=0.95]{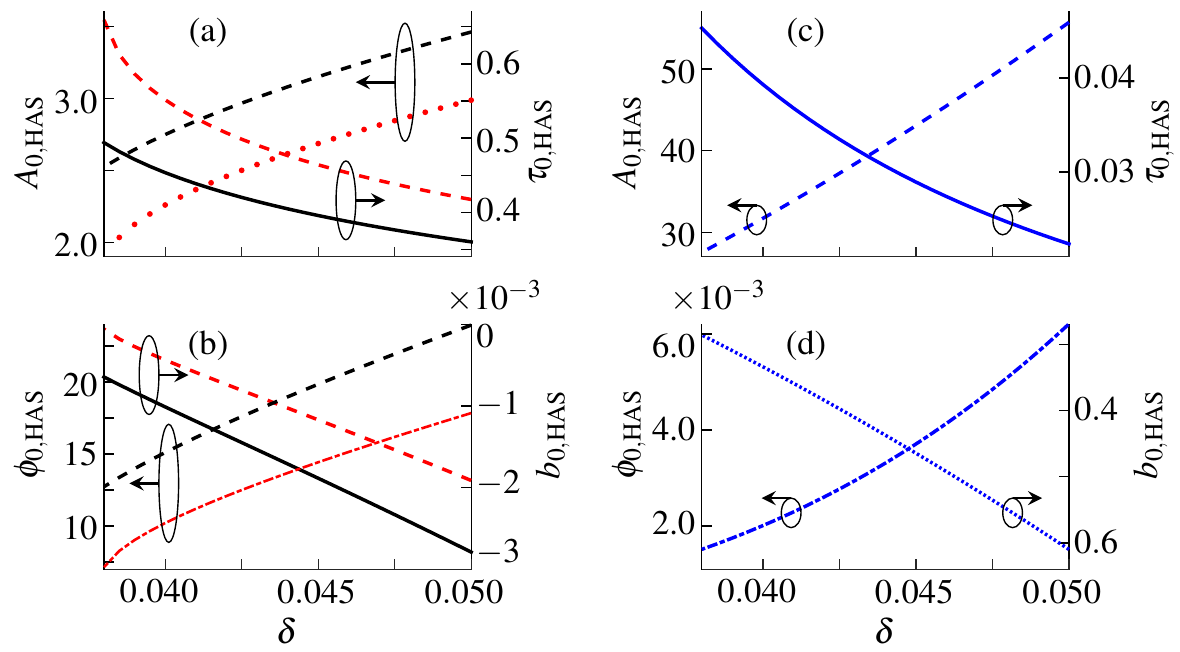} 
\vspace{-0.4cm}
\end{center}
\caption{With $\sigma=0.004$, the variation of the profile parameters of the high-amplitude solution (HAS): sub-figures (a) and (b) show the profiles of the cubic-quintic model and the sinusoidal model, while sub-figures (c) and (d) show the algebraic model. The black curves represent the pulse profiles with the cubic-quintic model, and, similarly, the blue curves represent the algebraic model, and the red curves represent the sinusoidal model. \label{fig:HAS-profile}}
\vspace{-0.2cm}
\end{figure}

In Fig.~\ref{fig:HAS-profile}, we show the variation of pulse profiles for the HAS with the three models when $\sigma=0.004$ and $\delta$ varies. 
As $\delta$ decreases, the peak amplitude $A_0$ and then the peak quintic loss decreases, and a saddle-node bifurcation eventually occurs when the quintic loss becomes insufficient to offset the nonlinear gain.
Here, the changes of the nonlinear loss and gain are dominated by the change of amplitude of the stationary pulse, which is different from the case of the LAS where the nonlinear terms are mainly impacted by the variation of the nonlinear coefficients $\sigma$ and $\delta$. 
Thus, for the HAS to remain stable, the pulse solution must have a sufficiently large peak amplitude. 
By comparing Figs.~\ref{fig:HAS-profile}(a) and~\ref{fig:HAS-profile}(c) with Fig.~\ref{fig:LAS-profile}(a), we find that for each of the three FSA models, the peak amplitude of the HAS is larger than that of the LAS, and hence the peak quintic loss from $\sigma|u|^4u$ is higher than that of the LAS, which
explains why the HAS remains stable at large values of $\delta$ when the LAS ceases to exist. 
We find in Figs.~\ref{fig:HAS-profile}(a) and~\ref{fig:HAS-profile}(c) that the cubic-quintic model generates pulses with higher amplitude and narrower width than does the sinusoidal model, while the algebraic model leads to pulse profiles that are significantly larger in amplitude and narrower in pulse width than for the other two FSA models. 
This behavior occurs because the nonlinear gain of the algebraic model increases monotonically with the input, while the nonlinear gain of the other two FSA models saturates when the input becomes sufficiently large, which is shown in Fig.~\ref{fig:absorbers}. 
Hence, as shown in Fig.~\ref{fig:allmodels}, with a fixed value of $\sigma$, the value of $\delta$ on $C_2$ is smallest for the algebraic model, next-smallest for cubic-quintic model, and is largest for the sinusoidal model.  

We find that the rate of phase rotation $\phi_0$ increases, but the chirp $b_0$ decreases, as $\delta$ increases, which is different from the profiles of the LAS, as can be seen comparing Fig.~\ref{fig:LAS-profile}(b) to Figs.~\ref{fig:HAS-profile}(b) and~\ref{fig:HAS-profile}(d). 
Both the rate of phase rotation and the chirp with the algebraic model are significantly larger than is the case for the other two models, as shown in Fig.~\ref{fig:HAS-profile}(d). 
For both the LAS and the HAS with any of the FSA models, we find that the rate of phase rotation $\phi_0$ grows together with the peak amplitude $A_0$, which implies that these stationary solutions are qualitatively similar to the soliton solution of the nonlinear Sch\"odinger equation~\cite{Drazin1989}.

\subsection{Stability of singular solutions of the GHME}

\subsubsection{Stability of the HAS as $\sigma\to0$\label{sec:asymptotic}}

When obtaining the stability structure in Figs.~\ref{fig:stability-cqme} and~\ref{fig:allmodels}, the boundary tracking algorithm is not able to proceed when $C_2$ approaches the $\delta$-axis where $\sigma=0$. 
We found that the HAS becomes increasingly narrower and more energetic as $\sigma\to0$, which implies that we are approaching a singular solution. 
By applying asymptotic perturbation theory to the GHME with the cubic-quintic FSA model~\cite{Wang:2016-PRA}, we proved that the HAS continues to exist and remains stable regardless of the magnitude of $\sigma$. 
Similarly, the HAS remains stable for both the sinusoidal model and the algebraic model as $\sigma\to0$ due to the similar dynamical structure that is visible in Fig.~\ref{fig:allmodels}. 

\subsubsection{Stability of the HAS as $\delta$ becomes sufficiently large}

\begin{figure}[!h]
	\begin{center}
		\includegraphics[scale=1]{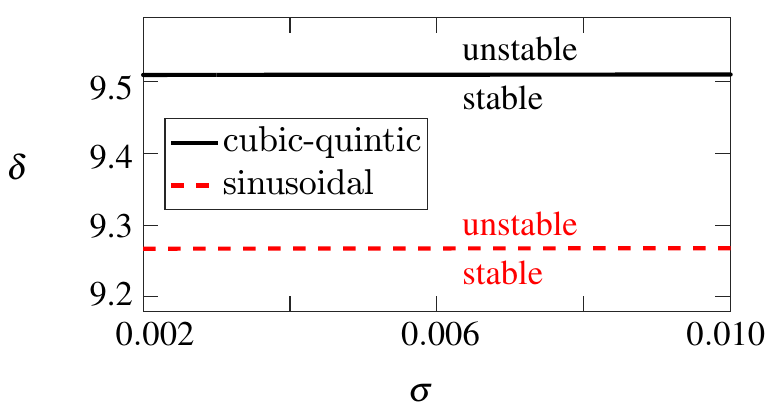} 
		\vspace{-1cm}
	\end{center}
	\vspace{0.6cm}
	\caption{The stability boundary of the GHME with the cubic-quintic model and the sinusoidal model at large values of $\delta$ as $\sigma$ varies. 
		The unstable region of both models lies above each curve. 
		The pulse solution of the GHME with the algebraic model is always stable when $\delta$ increases, as we prove in Appendix B.  \label{fig:boundary-edge}}
	\vspace{-0.2cm}
\end{figure}

With the cubic-quintic model, we found earlier that as $\delta$ increases, with $\sigma$ fixed, an edge bifurcation occurs in which two discrete modes bifurcate out of the continuous spectrum, which then become unstable via a Hopf bifurcation at $\delta\approx9.51$~\cite{Wang:2014}. 
Using the same approach, described in detail in Sec.~4.B.3 of~\cite{Wang:2014}, we find that the sinusoidal model exhibits the same qualitative behavior and becomes unstable when $\delta\approx9.26$, as shown in Fig.~\ref{fig:boundary-edge}. 
When the instability occurs, solution of the evolution equations shows that a shelf develops, which is a phenomenon that has been experimentally observed at high pump powers~\cite{Richardson1991}.
Because the gain turns over as $|u|$ increases, it becomes energetically favorable for the pulse duration to increase, rather than for its amplitude to increase.

With the algebraic model, the gain never turns over as $|u|$ increases, and a similar instability does not occur. 
In Fig.~\ref{fig:HAS-profile}, we show that the algebraic FSA model predicts stationary pulse solutions that are significantly shorter and more energetic than is the case for the other two FSA models. 
When $\delta>0.055$, we find that as $\delta$ increases with $\sigma$ fixed, the solution becomes increasingly narrow and energetic. 
Similar to what happens when $\delta$ is fixed and $\sigma\to0$, we find that the HAS approaches a singular solution, so that it becomes increasingly difficult to apply the boundary tracking algorithm. 
However, we can apply singular perturbation theory to show that a stationary solution exists at any $\delta$  and is always stable.
The detailed calculation is presented in Appendix B.

\section{Matching Experimental Results\label{sec:experiments}}

In Sec.~\ref{sec:stability} we showed that two stable stationary pulse solutions of the GHME coexist with an arbitrarily small value of $\sigma$, as long as $\sigma>0$, in contrast to the HME, where there is only one stable solution in a very limited region of the cubic nonlinearity $\delta$. 
Since any real system is likely to have a quintic component in its saturable absorber~\cite{Afanasjev:95}, the solutions of the GHME with higher nonlinearities should  provide a better approximation to the output pulses that have been observed in experiments than does the HME.
In this section, we will compare the stationary solutions that are predicted by the HME and the GHME. 

In Table~\ref{tab:params}, we show the parameters that we use in the comparative study, which are estimated based on the experiments. 
Set 1 of the parameters corresponds to a fiber laser with nonlinear polarization rotation~\cite{Washburn:04,newbury2005jqe}, and set 2 of the parameters corresponds to a Cr:LiSAF laser that is based on Kerr-lens modelocking~\cite{cihan2015}. 

\begin{table}[h!]
\begin{center}
\begin{tabu}{|c|c|c|c|c|c|c|c|c|c|}
\hline
\multicolumn{2}{|c|}{Parameter}
& $g_0$ & $l$ & $\gamma$ & $\omega_g$ & $\beta''$ & $T_RP_\mathrm{sat}$ & $\delta$ & $\sigma$ \\
\hline
\multirow{2}{*}{set 1} & value & $2.00$ & 1.65 & $4.10$ & $8.66$ & $-0.04$  & 0.30 & 0.87 & 0.55\\
 & unit & $1$ & 1 & kW$^{-1}$ & rad/ps & ps$^2$  & nJ & kW$^{-1}$ & kW$^{-2}$\\
\hline
\multirow{2}{*}{set 2} & value& $0.241$ & 0.045 & $0.65$ & $1257$ & $-8.0$  & 3.4 & 0.043 & 0.114\\
 & unit & $1$ & 1 & MW$^{-1}$ & rad/ps & $10^{-5}$ ps$^2$  & nJ & MW$^{-1}$ & MW$^{-2}$\\
\hline
\end{tabu}
\end{center}
\vspace{-0.3cm}
\caption{Values of parameters we use in validating the experimental results.\label{tab:params-exp}}
\vspace{-0.3cm}
\end{table}

\begin{figure}[!h]
\begin{center}
\includegraphics[scale=0.95]{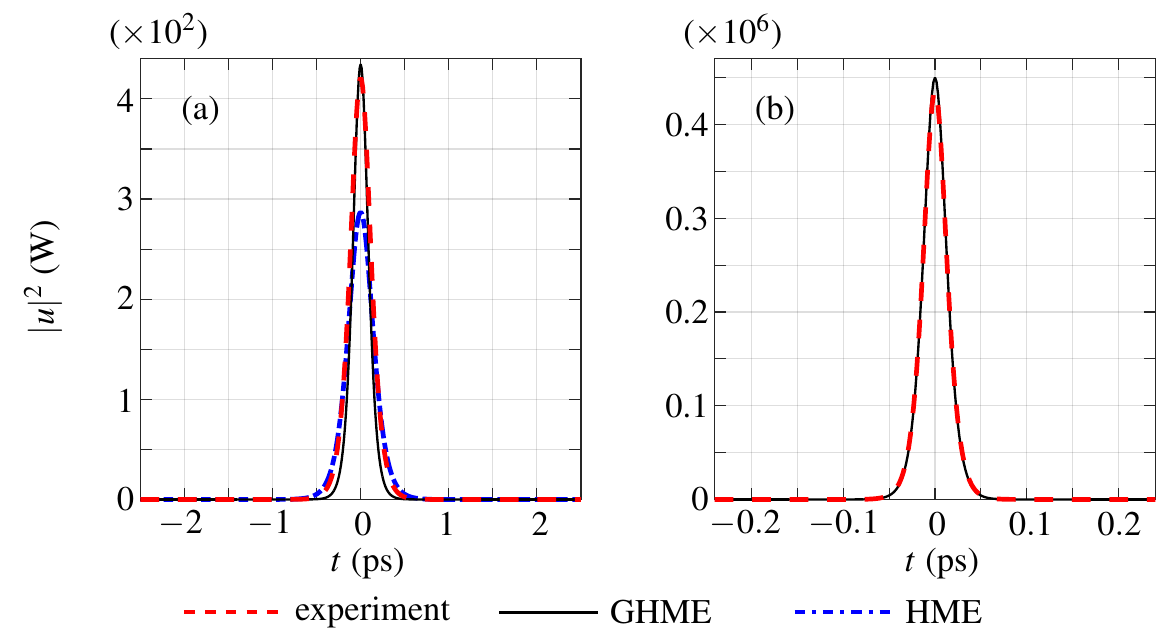} 
\vspace{-1.2cm}
\end{center}
\vspace{0.8cm}
\caption{Comparison of the computational stationary pulses with the experimental pulses using parameters in (a) set 1~\cite{Washburn:04,newbury2005jqe}, and (b) set 2~\cite{cihan2015}.  \label{fig:comp-expr}}
\vspace{-0.2cm}
\end{figure}

We show a comparison of the computational stationary pulses and the corresponding experimental results in Fig.~\ref{fig:comp-expr}. 
The fiber laser in~\cite{Washburn:04} generates a comb output with chirped pulses that have a duration of 210 fs and a peak power of 435 W. We show in Fig.~\ref{fig:comp-expr}(a) that using the GHME we are able to obtain a computational pulse with a full-width-half-maximum (FWHM) duration of $271$ fs and peak power of 421 W. 
We can achieve the closest match to the experimental pulse by setting $\delta=0.705\,\mathrm{kW}^{-1}$,  where the stationary pulse has an FWHM duration of 32.8 fs and a peak power of 287 W. 
When $\delta$ increases further, the HME solution ceases to exist.
The GHME solution is visibly a better match to the experimental pulse than is the HME solution. 
The modeling accuracy of the GHME could be further improved by a more accurate measurement of the parameters. 
In particular, the value of dispersion given in~\cite{Washburn:04} neglects the contribution from some cavity components. 

In Fig.~\ref{fig:comp-expr}(b), we show the comparison of the computational pulse with the experimental result corresponding to the solid state laser of~\cite{cihan2015}. 
In the experiment, a gain-matched output coupler is used to overcome the gain filtering effect. 
Using the transmission profile of the output coupler given in~\cite{cihan2015}, we are able to obtain accurately the pulse profile inside the laser cavity, where the pulse energy is 14.8 nJ and the FWHM width is 30 fs. 
We estimate the saturation power of the saturable absorption $P_\mathrm{ab}$ is 363 kW, and the saturable loss $f_0$ is 3\%. 
The system parameters are estimated following the approach in~\cite{Sander:09}. 
An algebraic model was used to model the FSA. By matching the cubic and the quintic coefficients, as seen from Fig.~\ref{fig:comp-expr}(b), we are able to obtain to a computational pulse of $14.9$ nJ with an FWHM width 29.2 fs, where the match is excellent. 
By comparison, no stable solution exists for the HME when we set $\sigma=0$.

\section{Summary}

We have compared three common models of FSA in the GHME to each other and to the HME using boundary tracking algorithms. 
These three FSA models are the cubic-quintic model, the sinusoidal model, and the algebraic model.
For all three models of the FSA, the stability region is greatly increased relative to the HME, in which the FSA only has a cubic nonlinearity.
The behaviors of these models are qualitatively similar, but quantitatively different, and the difference is more significant as the input power increases. 
At low pulse energies, any of these models can be used with an appropriate choice of parameters. 
At high pulse energies, the model must be carefully chosen to fit experimental systems, and it becomes questionable whether any averaged model can quantitatively reproduce the stability region of a real-world laser with lumped components. 
We will examine this question in future work.

The HME has at most only one stable stationary solution. 
By contrast, the GHME with any of the three FSA models can have two stable stationary solutions---an LAS that becomes equal to the stationary solution of the HME when $\sigma\to0$ and  an HAS that is qualitatively different. 
The LAS energy is dominated by a balance between the gain filtering, linear loss, and the cubic nonlinearity of the FSA while the HAS energy is dominated by a balance between the cubic and quintic nonlinearities of the FAS. 

The existence of the HAS is the key to understanding the greatly enhanced stability region that exists in the GHME and in experiments~\cite{Cundiff2003}. 
A quintic nonlinearity exists in the FSA of any real laser. 
Hence, we expect that the HAS will be present in any soliton laser. 
We also computed the predictions for the stationary solutions of the GHME and the HME to experiments. 
We found that the GHME provide much better agreement. 
We conclude that models of fast saturable absorption that include high-order nonlinearities should always be preferred to the HME when using averaged models. 

The existence and the stability of the HAS sheds further ligh on the possibility of finding high-energy and narrow pulses in soliton lasers. 
We show in~\cite{Wang:2016-PRA} that the HAS remains stable as $\sigma\to0$, while the pulse energy increases and the pulse duration decreases.
Hence, the GHME has a much larger region of stability than does the HME even when $\sigma$ is small. 
This result suggests that it may be possible to obtain stable modelocked pulses in experiments with high energy and large bandwidth if the quintic nonlinearity in the FSA can be decreased. 
In practice, this parameter is difficult to control, but our results provide motivation to make the attempt. 

\section*{Acknowledgments}
We thank Valentin Besse, Giuseppe D'Aguanno, and Thomas F. Carruthers for useful comments.  This work was supported by AMRDEC/DARPA, grant no. W31P4Q-14-1-0002.

\section*{Appendix A: Linear Stability of Pulse Solutions}

In the boundary tracking algorithms, we determine the stability of stationary solutions by calculating the eigenvalues of the linearized evolution equations. 
Soliton perturbation theory applied to the normalized nonlinear Schr\"odinger equation (NLSE) is a special case of this approach~\cite{Kapitula:98,kaup1990}. 
We may write the NLSE as
\begin{align}\label{eq:nls}
\pypx{u}{z}&=i\phi u + \frac{i}{2}\pynpx{u}{t}{2} + i|u|^2u,
\end{align}
which has the stationary soliton solution
\begin{align}
u_0(t)=\sech(t), \quad \phi_0 = 1/2.
\end{align}
Linearization of the NLSE leads to
\begin{equation}\label{eq:nls-linearize}
\begin{aligned}
\pypx{\Delta u}{z} &= i\phi_0 \Delta u  + \frac{i}{2}\pynpx{\Delta u}{t}{2} + (2i\sech^2 t) \Delta u + (i\sech^2 t) \Delta \bar{u},  \\
\pypx{\Delta \bar{u}}{z} &= -i\phi_0 \Delta  \bar{u} - \frac{i}{2}\pynpx{\Delta  \bar{u}}{t}{2} - (2i\sech^2 t) \Delta  \bar{u} - (i\sech^2 t) \Delta {u}, 
\end{aligned}
\end{equation}
where $u = u_0+\Delta u$, $\bar{u}=u_0^*+\Delta \bar{u}$, and $[\Delta u,\Delta \bar{u}]^T$ is a perturbation to the stationary pulse $[u_0(t),u_0^*(t)]^T$, and the superscript $T$ indicates the transpose. 
We next formulate the eigenvalue problem
\begin{align}
\pypx{\Delta \vec{u}}{z}&={\mathcal{L}}(u_0)\Delta \vec{u} = \lambda\Delta \vec{u}, \end{align}
where $\Delta\vec{u}=[\Delta u, \Delta\bar{u}]^T$, and $\mathcal{L}$ is matrix form of the right hand side of \quoeq{eq:nls-linearize}. 
Here, we suppose that $\Delta \vec{u}$ and $\lambda$ are an eigenvector and its corresponding eigenvalue.
If any eigenvalues have positive real parts, the pulse solution is unstable. 
A more detailed discussion based on the GHME with the cubic-quintic FSA model can be found in Sec.~3 of~\cite{Menyuk:2016}, as well as Sec.~3 of~\cite{Wang:2014}---in which the stationary solutions are referred to as equilibrium solutions. 

\begin{figure}[!h]
	\begin{center}
		\includegraphics[scale=1]{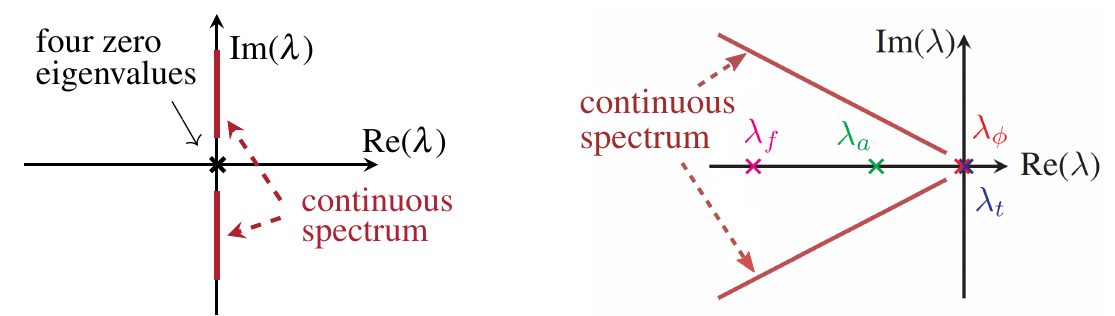} 
		\caption{Illustrations of the linearized spectrum of the eigenvalues of (a) the nonlinear Schr\"odinger equation and (b) the generalized Haus modelocking equation.\label{fig:spectrum}}
	\end{center}
	\vspace{-0.5cm}
\end{figure}

In Fig.~\ref{fig:spectrum}, we illustrate the eigenvalues' distribution on the complex plane, which we refer to as the (linearized) spectrum, for both the NLSE and the GHME. 
In both cases, the spectrum includes two branches of the continuous spectrum that are symmetric about the real axis and four real discrete eigenvalues that correspond physically to perturbations of the stationary solution\rq{}s central time ($\lambda_t$), central phase ($\lambda_\phi$), amplitude ($\lambda_a$), and central frequency ($\lambda_f$)~\cite{haus206583,kaup1990}. 
Any positive real part of the spectrum indicates instability. 

As shown in Fig.~\ref{fig:spectrum}(a), for the NLSE, all four discrete eigenvalues $\lambda_t$, $\lambda_\phi$, $\lambda_f$, and $\lambda_a$ equal zero, and the two branches of the continuous spectrum are aligned along the imaginary axis. 
Hence, the NLSE soliton is neutrally stable, which means that perturbations of the stationary soliton will not decay as $z$ increases. 

By comparison, when a pulse solution of the GHME is linearly stable, the real parts of all eigenvalues are negative except for $\lambda_t$ and $\lambda_\phi$, which equal $0$, as shown in Fig.~\ref{fig:spectrum}(b). 
These eigenvalues must equal zero because the GHME, as well as its linearization, are invariant with respect to time and phase shifts. 
In the case of Fig.~\ref{fig:spectrum}(b), the spectrum indicates that any perturbation to the pulse solution---expect for those that are proportional to the eigenfunctions of the phase shift or the time shift---will decay as the pulse propagates. 
The pulses's central phase and central time undergoes a random walk in the presence of noise. 
In modern comb systems, one uses electronic feedback to force a phase or a time shift to decay exponentially toward a reference phase and time~\cite{Menyuk:07,Cundiff2003,Jones635,cundiff2005}.
From a theoretical standpoint, the electronic feedback control moves the two eigenvalues $\lambda_t$ and $\lambda_\phi$ to the left in Fig.~\ref{fig:spectrum}. 
This effect has been experimentally observed~\cite{Wahlstrand:07}.

\section*{Appendix B: Stability of GHME with the algebraic FSA Model}
The GHME with the algebraic model can be written as
\begin{equation}\label{eq:hme-al}
\begin{aligned}\displaystyle
\pypx{u}{z}=\bigg[-i\phi-\frac{l}{2}+\frac{g(\abs{u})}{2} \bigg(1+\frac{1}{2\omega_g^2}\pynpx{}{t}{2}\bigg) - \frac{i\beta''}{2}\pynpx{}{t}{2} + i\gamma|u|^2  + \frac{\delta|u|^2}{1+\sigma/\delta |u|^2}\bigg]u,
\end{aligned}
\end{equation}
where we show stable operation region in Fig.~\ref{fig:allmodels} and the variation of the stationary pulse profile of the HAS in Figs.~\ref{fig:HAS-profile}(c) and~\ref{fig:HAS-profile}(d). 
As $\delta$ further increases beyond $\sim0.055$, the pulse profile becomes increasingly narrow and energetic, and it becomes computationally difficult to obtain the stationary solution accurately.  

We also find that, as $\delta$ increases, the behavior of the HAS that we have computationally obtained becomes increasingly similar to a soliton solution
\begin{align}\label{eq:nls-soliton}
u_0 = A_0\,\sech\paren{\sqrt{-\frac{\gamma}{\beta}}A_0\,t},\quad \phi_0=\frac{\gamma }{2}A_0^2
\end{align}
of the NLSE  
\begin{align}\label{eq:nls-terms}
\pypx{u}{z}=-i\phi u - \frac{i\beta''}{2}\pynpx{u}{t}{2} + i\gamma|u|^2u.
\end{align}
The NLSE terms in \quoeq{eq:hme-al} increasingly dominate the stationary pulse, so that the remaining terms that govern gain and loss can be treated perturbatively~\cite{Kodama1992,Gupta1978}. 

We observe that, as $\delta$ grows, the HAS becomes increasingly narrow and energetic, and thus $f_\mathrm{sa,al}$, given by \quoeq{eq:algebraic}, becomes increasingly similar to a linear gain,
\begin{align}
f_\mathrm{sa,al}(|u|) = \frac{\delta |u|^2}{1+\paren{\sigma/\delta}|u|^2} \approx \frac{\delta^2}{\sigma}.
\end{align}
It is for this reason that the HAS approaches the NLSE soliton asymptotically when $\delta$ is large. 
As discussed in Appendix A, the stationary pulse is invariant to perturbation modes that correspond to phase rotation and time translation, and the stationary pulse is linearly stable with respect to frequency perturbations due to the frequency filter term $[{g(|u|)}/(4\omega_g^2)] \partial^2 u/\partial t^2$. 
Thus, we always find $\lambda_t=\lambda_\phi=0$ and that $\lambda_f$ as well as the continuous spectrum have negative real parts. 
Therefore, in order to prove the stability of the HAS, we need only show that the amplitude eigenvalue $\lambda_a$ remains negative as $\delta$ increases. 

Here, we formulate a reduced equation based on the GHME in \quoeq{eq:hme-al}~\cite{Kapitula:98,kaup1990,Menyuk:2016}. 
We multiply both sides of \quoeq{eq:hme-al} by $u^*$ and obtain
\begin{align}\label{eq:to-integration}
u^*\pypx{u}{z} = \sqpr{-i\phi -\frac{l}{2}+\frac{g(\abs{u})}{2}}|u|^2 +\sqpr{\frac{g(\abs{u})}{4\omega_g^2} - \frac{i\beta''}{2}}u^*\pynpx{u}{t}{2}  + i\gamma|u|^4 + \frac{\delta|u|^4}{1+\sigma/\delta |u|^2}.
\end{align}
Then we take the complex conjugate of \quoeq{eq:to-integration2},
\begin{align}\label{eq:to-integration2}
u\pypx{u^*}{z} = \sqpr{i\phi -\frac{l}{2}+\frac{g(\abs{u})}{2}}|u|^2 +\sqpr{\frac{g(\abs{u})}{4\omega_g^2} + \frac{i\beta''}{2}}u\pynpx{u^*}{t}{2}  - i\gamma|u|^4 + \frac{\delta|u|^4}{1+\sigma/\delta |u|^2}.
\end{align}
We now add Eqs.~(\ref{eq:to-integration}) and~(\ref{eq:to-integration2}), and we obtain
\begin{align}\label{eq:preintegration}
\pypx{|u|^2}{z} = \sqpr{g(|u|)-l}|u|^2
+\frac{g(|u|)}{4\omega_g^2} \paren{u^*\pynpx{u}{t}{2} +u\pynpx{u^*}{t}{2}} 
+ \frac{i\beta''}{2}\paren{u\pynpx{u^*}{t}{2}-u^*\pynpx{u}{t}{2}}
+ \frac{2\delta|u|^4}{1+\sigma/\delta |u|^2},
\end{align}
where $g(|u|)={g_0}/\paren{1+w/E_\mathrm{sat}}$ and $w = \int_{-T/2}^{T/2}|u|^2\dint t$. By integrating both sides of \quoeq{eq:preintegration} in $t$ we obtain
\begin{align}\label{eq:integrated}
\pypx{w}{z} = \sqpr{g(|u|)-l}w - \frac{g(|u|)}{2\omega_g^2}\int_{-T/2}^{T/2}\left|\pypx{u}{t}\right|^2 \dint t + \int_{-T/2}^{T/2}\frac{2\delta|u|^4}{1+\sigma/\delta |u|^2}\dint t.
\end{align} 

Since the HAS approaches the NLSE soliton as $\delta$ increases, we substitute \quoeq{eq:nls-soliton} into \quoeq{eq:integrated} and obtain an ordinary differential equation for the amplitude $A_0$
\begin{equation}
\begin{aligned}\label{eq:A0ODE}
\dydx{A_0}{z}=f(A_0)= &\,  \frac{g_0{E_\mathrm{sat}}}{{E_\mathrm{sat}} + {2A_0}\sqrt{-{\beta''}/{\gamma}}} \paren{A_0+\frac{\gamma A_0^3}{6\beta''\omega^2}}  -lA_0 \\
& + \frac{2\delta^2A_0}{\sigma} - \frac{\delta^3\sigma^{-3/2}}{\sqrt{\delta+\sigma A_0^2}}
\log\paren{\frac{\sqrt{\delta+{\sigma} A_0^2}+\sqrt{\sigma} A_0}{\sqrt{\delta+\sigma A_0^2}-\sqrt{\sigma} A_0}}.
\end{aligned}
\end{equation}
The amplitude $A_0$ of a stationary solution $u_0$ satisfies $\dint A_0/\dint z=0$. 
If there is a perturbation of the amplitude $A = A_0 + \Delta A$, we can then linearize \quoeq{eq:A0ODE} and obtain
\begin{align}
\dydx{\Delta A}{z}\approx \dydx{f(A)}{A}\bigg|_{A=A_0}\Delta A = \lambda_a \Delta A,
\end{align}
where
\begin{equation}
\begin{aligned}
\lambda_a = &\, \frac{g_0{E_\mathrm{sat}}}{{E_\mathrm{sat}} + {2A_0}\sqrt{-{\beta''}/{\gamma}}}  \paren{1+\frac{\gamma A_0^2}{2\beta\omega_g^2}} 
- \frac{2g_0 E_\mathrm{sat} \sqrt{-\beta/\gamma}}{\paren{E_\mathrm{sat} + {2A_0}\sqrt{-{\beta''}/{\gamma}}}^2} \paren{A_0+\frac{\gamma A_0^3}{6\beta\omega_g^2}} \\
& -l + \frac{2\delta^2}{\sigma}  
+ \frac{\delta^3 A_0}{\sqrt{\sigma}(\delta+\sigma A_0^2)^{3/2}}\log\paren{\frac{\sqrt{\delta+{\sigma} A_0^2}+\sqrt{\sigma} A_0}{\sqrt{\delta+\sigma A_0^2}-\sqrt{\sigma} A_0}} 
- \frac{2\delta^3}{\sigma(\delta+\sigma A_0^2)}.
\end{aligned}
\end{equation}

\begin{figure}[!h]
\begin{center}
\includegraphics[scale=1]{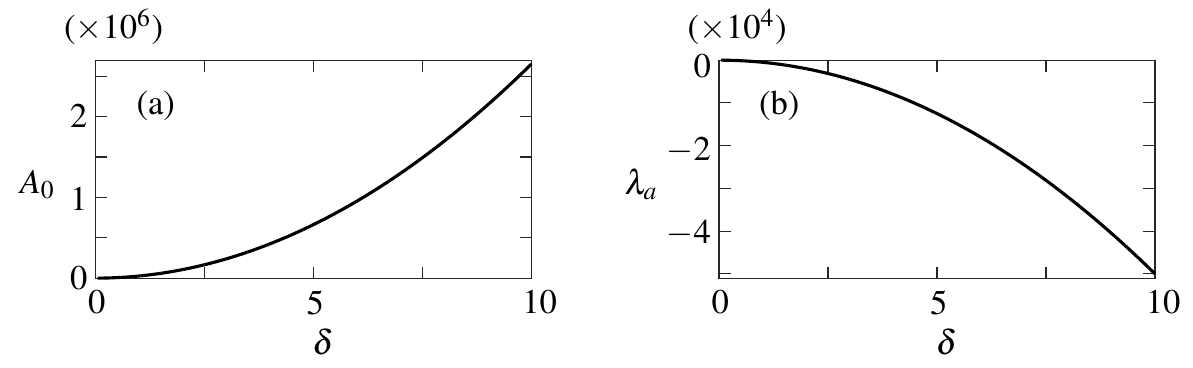} 
\vspace{-0.5cm}
\end{center}
\caption{As the nonlinear gain coefficient $\delta$ increases, for the GHME in \quoeq{eq:hme-al}, the variations of (a) the peak amplitude of the HAS $A_0$ and (b) the amplitude value $\lambda_a$.  \label{fig:A-stability}}
\vspace{-0.2cm}
\end{figure}

We now solve \quoeq{eq:A0ODE} computationally and evaluate $\lambda_a$. 
We show the results in Fig.~\ref{fig:A-stability}. 
As $\delta$ increases, we observe that $A_0$ increases while $\lambda_a$ becomes more negative. 
Hence, the stationary solution $u_0$ is stable against perturbations that are proportional to the amplitude eigenfunction. 
Considering the asymptotic behavior of both the pulse profile and the nonlinear gain of $f_\mathrm{sa}(|u|)$, we expect $\lambda_a$ to become more negative as $\delta$ further increases. 
Therefore, all eigenvalues are negative except $\lambda_t=\lambda_\phi=0$ as $\delta$ increases.
This result indicates that, for the GHME with the algebraic FSA model, the HAS remains stable for large values of $\delta$, and its stable region is only bounded below by the Hopf bifurcation limit, which is indicated by $C_2$ in Fig.~\ref{fig:allmodels}.

\end{document}